

\input phyzzx.tex

\def\d {\partial}
\def\dd {\bar\partial}

\def\a {\alpha}

\def\h {\hat g}
\def\t {\theta}
\def\f {\phi}
\def\b {\beta}

\titlepage
Dipartimento di Fisica\par
Universit\`a di Roma "La Sapienza"\par
I.N.F.N. -- Sezione di Roma\par\par
{\bf Preprint n.852}\par
hepth@xxx/9201068 \par

\title{On the Black-Hole Conformal Field Theory Coupled to the Polyakov's
String Theory. A Non Perturbative Analysis}

\author {M. Martellini\foot{On leave of absence from Dipartimento di Fisica,
Universit\'a
di Milano, Milano, Italy and I.N.F.N., Sezione di Pavia, Italy}}
\address{I.N.F.N., Sezione di Roma "La Sapienza", Roma, Italy}
\author {M. Spreafico}
\address{Dipartimento di Fisica, Universit\'a di Milano, Milano, Italy}
\author {K. Yoshida}
\address{Dipartimento di Fisica, Universit\'a di Roma "La Sapienza", Roma,
Italy
and I.N.F.N., Sezione di Roma, Italy}

\abstract
We couple the 2D black-hole conformal field theory discovered by Witten
to a $D-1$ dimensional
Euclidean bosonic string. We demonstrate that the resulting planar
(=zero genus) string susceptibility
is real for any $0\leq D \leq 4$.
\endpage

\pagenumber=1

Witten [1] has constructed a modular invariant SL(2,$\Re$)/U(1) coset
model as a gauged
SL(2,$\Re$)-Wess-Zumino-Witten (WZW) theory which, for a large Kac-Moody
level $k$, describes
a bosonic string propagating over a two dimensional black-hole target
space-time.
If we denote the two target space-time coordinates by ($r,\t$),
one sees that for large $r$ (radial component),
the same $r$ can be identified with the Liouville field
and one gets the two dimensional space-time that appears
in the standard $c=1$ non-critical string theory.

However, for finite $r$, the two theories are different.
In fact the exact black-hole quantum field theory seems
in some sense more "fundamental". Indeed, quoting
Witten [1], "the region $k>{9\over 4}$, which corresponds to what
is normally regarded as the forbidden region of $c>1$
in Liouville theory, makes perfect sense for Euclidian black-hole."

Therefore, replacing the standard Liouville theory with the
Euclidian black-hole, one could get
past the Liouville theory barrier at $c=1$.

In this letter, continuing the line of Witten's arguments
starting from these observations,
we would like to show that if one couples the conformal
field theory (CFT) governing
the Euclidean 2D black-hole to $D-1$ massless scalar free
fields described by a Polyakov's
action with the central charge $c$ (before coupled to
the black-hole), then one gets a well defined theory
also in the physically interesting range $c=D>1$.
The original Witten's theory corresponds to the case $D=1$.

Our starting point is the result of Bershadsky-Kutasov [2]
which shows that the Euclidian
SL(2,$\Re$)/U(1) CFT with the action $S_W$ in free field
representation is identical quantum mechanically
to a Liouville field $\f$ coupled to a $c=1$ conformal
field $X^{\scriptscriptstyle 0}$, with a \underbar {non standard} cosmological
constant term O($\mu$). Indeed, here the cosmological
constant $\mu$ is \underbar{not} controlling the area
(identity) operator $\hat A = \sqrt{\h}e^{\beta\f}$
(the physical world-sheet metric being taken as
$g_{\mu\nu} =e^{\b\f} \h_{\mu\nu}$), but rather the
primary spinless operator $\Phi$ of lower dimension
$\Delta_0$ of the SL(2,$\Re$)/U(1) theory before the
"gravitational dressing" by $\sqrt{\h} e^{\beta\f}$.

Explicitly the action $S_W$ can be rewritten as
$$
S_W = {1\over 8\pi}\int d^2 z \sqrt{\h}\h^{\a\b}
[\d_\a\f\d_\b\f +\d_\a X^{\scriptscriptstyle 0}\d_\b X^{\scriptscriptstyle 0}]
-{Q\over 4}\int d^2 z \sqrt{\h} \hat R^2(\h)\f
+\mu\int d^2 z \sqrt{\h} e^{\b\f}\Phi(\f,X^{\scriptscriptstyle 0})
\eqn\uno
$$
where
$$
\Phi(\f,X^{\scriptscriptstyle 0}) = (a\d\f+ib\d X^{\scriptscriptstyle 0})
(a\dd\f+ib\dd X^{\scriptscriptstyle 0})
\eqn\due
$$
and (a,b) are the two free parameters fixed by
the conformal invariance, and whose
explicit values are not relevant for our purpose.

The total action is then obtained by adding to
\uno\ the action of reparametrization ghosts
plus the Polyakov's string action for $D-1$ massless scalar fields
$$
S_P = {1\over 8\pi}\int d^2 z \sqrt{\h} \h^{\a\b}\sum_{a=1}^{D-1}
(\d_\a X^a\d_\b X^a)
\eqn\tre
$$
Returning to eq. \uno\ , we see that the cosmological
constant operator $\sqrt{\h}e^{\b\f}\Phi$ is
a conformal field of dimension (1,1) under conformal
transformation, if the coefficient $\b$ satisfies the
constraint equation
$$
\Delta_0 - {1\over 2} \b (Q+\b) = 1
\eqn\quattro
$$
where $\Delta_0$ is the dimension of the primary
spinless field $\Phi$, eq. \due\ .

In eq. \quattro\ , $Q$ is the background charge
whose value is fixed by vanishing of the
total charge.
$$
c_{\scriptscriptstyle{\f}} + c_{\scriptscriptstyle{X^{\scriptscriptstyle 0}}}
+ c_{\scriptscriptstyle{\{ X^a \} }} + c_{\scriptscriptstyle{gh}} =0
\eqn\cinquea
$$
where
$$
\left \{
\eqalign{
&c_{\scriptscriptstyle\f} = 1+3Q^2 \cr
&c_{\scriptscriptstyle{X^{\scriptscriptstyle 0}}} = 1 \cr
&c_{\scriptscriptstyle{\{ X^a \} }} = D-1 \cr
&c_{\scriptscriptstyle{gh}} = -26 \cr}
\right.
\eqn\cinqueb
$$
Thus one finds from eq. \cinquea\ and \cinqueb\ that
$$
Q=\sqrt{{25-D \over 3}}
\eqn\sei
$$
Substituting this value for $Q$ into eq. \quattro\ we get
$$
\b_{\pm}(\Delta_0)=-{1\over 2\sqrt{3}}(\sqrt{25-D}\pm\sqrt{1-D+24\Delta_0})
\eqn\sette
$$
We shall choose in what follows the plus sign in \sette\ , i.e. $\b=\b_+$.

Let us then assume that the by now classical ansatz
of Ref.[4] can be applied to our present
situation.
The calculation of the string susceptibility $\gamma_h$
for power-like scaling of the total partition function
$Z(\mu)$ on a closed compact surface of genus $h$
follows in our case from the \underbar{constant} shift in the
Liouville field $\f$ (cf. Ref.[4])
$$
\f \to \f+{ln(\mu)\over \b_+ (\Delta_0)}
\eqn\otto
$$
where $\b_+ (\Delta_0)$ has replaced the usual $\b_+ (0)$.

In this manner, one obtains an "effective" string susceptibility as
$$
\eqalign{
&Z(\mu)\sim \mu^{\gamma_h^{eff} -3}\cr
&\gamma_h^{eff} = {\chi(h) Q\over 2\b_+(\Delta_0)}+2\cr}
\eqn\nove
$$
and, in particular for $h=0$
$$
\gamma_0^{eff} = {1\over 12} {D-1-24\Delta_0 +
\sqrt{(25-D)(1-D+24\Delta_0)}\over 1-\Delta_0}
\eqn\dieci
$$
Here $\chi(h)=2-2h$ is the Euler characteristic.

It is also possible to recast these discussions
in terms of the U(1) gauged $\hat{SL_k}(2,\Re)$
Kac-Moody algebra.

The level $k$ is now renormalized according to the
relation \cinquea\ ,where $c_{\f}=2+{6\over k-2}$.
One gets
$$
(2+{6\over k-2})+(D-1)-26=0
$$
from which it follows
$$
k=2+{6\over 24-\tilde D}
$$
where we write $\tilde D$ for $D-1$.
The Kac-Moody (KM) conformal dimension $\Delta_{KM}$
of a primary coset operator
$\Phi_{l,m,n}$ of the Euclidean black-hole CFT
characterized by the integer $m, n$
($m$($n$) is to be interpreted as the discrete momentum
(winding number) of $X^{\scriptscriptstyle 0}$) and
by the SL(2,$\Re$) isospin $l$, is given in terms of
the principal discrete series
of the unitary representation of $\hat{SL_k}(2,\Re)$ [5].
$$
\Delta_{l,m,n}\equiv\Delta_{KM}(\Phi_{l,m,n})=-{l(l+1)\over k-2}+
{(m+nk)^2\over 4k}
\eqn\dodici
$$
The fusion algebra implies the constraints
$$
\left \{
\eqalign{
&|n-k|\ge|m|\cr
&l=r-{1\over 2}|nk|+{1\over 2}|m|<-{1\over 2}\cr}
\right.
\eqn\tredici
$$
where $r$ runs over the non-negative integers.

In our scenario, the gravitational dressing of
a (spinless) primary field $v$ of bare
dimension $\Delta_B$, is realized by the operator $\Phi_{l,m,n}$ with
$m=n=0$, i.e. \underbar{without}
the $S^1$-compactified field $X^{\scriptscriptstyle 0}$.
In other words, $\Phi_l \equiv\Phi_{l,0,0}$
mimics the Liouville field operator $exp(\b\f)$.
Thus the constraint $\Delta_B+\Delta_l=1$,
where $\Delta_l\equiv\Delta_{l,0,0}$, gives
$$
\left \{
\eqalign{
&\Delta_B(v)=1-\Delta_l=1+{l(l+1)\over k-2}\cr
&l= -1, -{3\over 2}, ...\cr}
\right.
\eqn\quattordici
$$
Thus the lowest bare dimension $\Delta_0=\Delta_{B,mim}(v)$
is reached by eq. \quattordici\
at $l=-1$, namely $\Delta_0 =1$.
Applied to the cosmological term operator $e^{\b\f}\Phi$ in eq.
\uno\ (i.e. $v\equiv\Phi$), above general
argument implies that $\Delta_0(\Phi)=1$.

Such a result is in agreement with the free field realization
of the SL(2,$\Re$)/U(1) coset theory.
Indeed the operator $\Phi$ defined by eq. \due\ may be rewritten as [2]
$$
\Phi\sim J_{\zeta}^{tot}\overline J_{\zeta}^{tot}
\eqn\quindici
$$
where $J_{\zeta}^{tot}$ is the total axial U(1) current.

But $J_{\zeta}^{tot}(\overline J_{\zeta}^{tot})$ is a primary
field of dimension (1,0) ((0,1)), and hence
$\Phi$ is a primary (spinless) field of dimension
$$
\Delta_0 = \overline {\Delta}_0 = 1
$$

As a consequence of this fact, we see that the planar
(i.e. zero genus) effective
string susceptibility $\gamma_0^{eff}\equiv\gamma_{h=0}^{eff}$
given by eq. \nove\ has the value
$$
\gamma_0^{eff} = {Q\over \b_+(1)}+2=-{Q\over Q}+2=1
\eqn\sedici
$$
since $\chi(h=0)=2$ and $\b_+(1)=-\sqrt{{25-D\over 3}}= -Q$.

This last result does not depend on $D$ and, if $h\not= 0$,
is universal, i.e. it is a
function only of the genus of the string world-sheet.

It is to be understood that at the level of the partition
function, $\gamma\equiv\gamma_0^{eff}> 0$
is the only quantity which indicates the presence of $c>1$
matter coupled to world-sheet,
and there is numerical evidence that for $c>1$ one has that $\gamma>1$ [6].

\endpage

ACKNOWLEDGEMENT

One of the authors (M.M.) have benefited from very useful
discussions with M. Bianchi.

REFERENCES

\parindent=0.pt

[1] E. Witten, Phys. Rev. D44, 314 (1991)

[2] M. Bershadsky and D. Kutasov, Phys. Lett. B266, 345 (1991)

[3] M. Wakimoto, Comm. Math. Phys. 104, 605 (1986)

[4] J. Distler and H. Kawai, Nucl. Phys. B321,509 (1989)

[5] R. Dijkgraaf, E. Verlinde and H. Verlinde, PUPT-1252 and
IASSNS-HEP-91/22 preprint (1991)

[6] J. Ambjorn, B. Durhams and J. Frohlich, Nucl. Phys. B257 [FS 14],
3 (1985), B275 [FS 17], 161 (1986);
with P. Orland, Nucl. Phys. B270 [FS 16], 475 (1986)

\end